# Electronic transport in polycrystalline graphene

Oleg V. Yazyev[1,2] & Steven G. Louie[1,2]

[1]*Department of Physics, University of California, Berkeley, California 94720, USA*

[2]*Materials Sciences Division, Lawrence Berkeley National Laboratory, Berkeley, California 94720, USA*

**Most materials in available macroscopic quantities are polycrystalline. Graphene, a recently discovered two-dimensional form of carbon with strong potential for replacing silicon in future electronics,[1–3] is no exception. There is growing evidence of the polycrystalline nature of graphene samples obtained using various techniques.[4–13] Grain boundaries, intrinsic topological defects of polycrystalline materials,[14] are expected to dramatically alter the electronic transport in graphene. Here, we develop a theory of charge carrier transmission through grain boundaries composed of a periodic array of dislocations in graphene based on the momentum conservation principle. Depending on the grain boundary structure we find two distinct transport behaviours – either high transparency, or perfect reflection of charge carriers over remarkably large energy ranges. First-principles quantum transport calculations are used to verify and further investigate this striking behaviour. Our study sheds light on the transport properties of large-area graphene samples. Furthermore, purposeful engineering of periodic grain boundaries with tunable transport gaps would allow for controlling charge currents without the need of introducing bulk band gaps in otherwise semimetallic graphene. The proposed approach can be regarded as a means towards building practical graphene electronics.**

Graphite, a precursor in producing graphene by exfoliation, exists at large scale in polycrystalline form only. Tilt grain boundaries in graphite have been extensively investigated over the past twenty years.[4–8] Epitaxial growth of graphene on a variety of substrates is envisaged to become a primary route towards industrial scale production of graphene-based electronic devices. Again, substrate imperfections and kinetic factors of the growth process inevitably result in grain boundary defects (for examples see Refs. 9–12). On the other hand, controlled manufacturing of structurally well-defined line defects in epitaxial graphene has been demonstrated very recently.[13] Scanning probe microscopy studies showed that such one-dimensional defects introduce pronounced perturbation into the electronic structure.[7–8,13] Importantly, grain boundaries in graphene appear as well-ordered periodic structures with typical periodicities of 1–5 nm. This suggests that crystal momentum conservation would play a crucial role in the transmission of charge carriers across these topological defects.

In two-dimensional (2D) materials such as graphene, grain boundaries are the one-dimensional (1D) interfaces between two domains of material with different crystallographic orientations. The structure of grain boundaries is defined by $\theta_L$ and $\theta_R$, the angles between the corresponding crystallographic directions in the two domains and the normal of the boundary line (see Fig. 1a.). In our consideration, we adopt an approximation which models grain boundaries as periodic arrays of dislocations.[14,15] Their periodicity is defined by the commensurability condition, either exact or close matching, of two translation vectors ($n_L$, $m_L$) and ($n_R$, $m_R$) belonging to the left and right crystalline domains, respectively, and oriented along the boundary line. This construction is illustrated on a representative example of a grain boundary structure shown in Figure 1a. The repeat vector $\vec{d}$ of this model structure is defined by the (5,3) and (7,0) matching vectors; the length of both vectors is $d_{(n,m)} = |n\,\vec{a_1} + m\,\vec{a_2}| = a_0\sqrt{n^2 + nm + m^2} = 1.72$ nm ($a_0 = 0.246$ nm is the length of unit vectors $\vec{a_1}$ and $\vec{a_2}$ of the graphene lattice). Lowest-energy atomic structures of the interface regions of grain boundaries are constructed from pentagon-hexagon pairs, elementary dislocations in graphene, as we have recently shown in Ref. 16. A possible atomic structure of the (5,3)|(7,0) grain boundary accommodates three such dislocations per 1D unit cell (Fig. 1a).

The low-energy charge carriers in graphene are characterized by the linear dispersion $E(\vec{k}) = \hbar v_F |\vec{k}|$, with Fermi velocity $v_F \approx 10^6$ m/s, and momentum $\vec{k}$ measured relatively to the Dirac points, the corners (points $K$ and $K'$) of the hexagonal 2D Brillouin zone (BZ). When crossing a grain boundary, the charge carriers experience an effective rotation of the Brillouin zone as shown in Figure 1b. From translational symmetry, elastic transmission implies that both energy $E$ and momentum $k_\parallel$ parallel to the interface are conserved. The correspondence between momenta $\vec{k}$ in the 2D BZ of graphene and $k_\parallel \in [-\pi/d; \pi/d)$, the 1D mini-BZ defined by the matching vector $(n, m)$ of the periodic boundary, is made by folding momentum space along the lines $k_\parallel = (2n+1)\pi/d$ ($n \in \mathbb{Z}$) and projecting the states on the 1D mini-BZ, as illustrated in Figure 2a. Such geometric construction is similar to the one used for explaining the `multiple-of-3' trends in the electronic structure of carbon nanotubes.[17,18] Two cases are possible: If $n - m = 3q$ ($q \in \mathbb{Z}$), both Dirac points $K$ and $K'$ are intersected by the solid lines defined by $k_\parallel = 2n\pi/d$ ($n \in \mathbb{Z}$), and, hence, are mapped onto the $\Gamma$ point ($k_\parallel = 0$) in the 1D mini-BZ (Fig. 2b). Otherwise, the distance to the closest solid line is $\Delta k_\parallel = 2\pi/(3d)$, i.e., the two Dirac points are folded separately onto $k_\parallel = -2\pi/(3d)$ and $k_\parallel = 2\pi/(3d)$. If both matching vectors $(n_L, m_L)$ and $(n_R, m_R)$ correspond to the same case, either $n - m = 3q$ or $n - m \neq 3q$, no suppression of transmission by mismatch of $k_\parallel$ (selection by momentum) takes place. For a conductance channel with given $k_\parallel$ and $E$ on one side of grain boundary, there always exists a channel characterized by the same momentum and energy on the other side (Fig. 2c). Below, these two situations will be referred to as class Ia and class Ib grain boundaries, respectively. However, in the case with $n_L - m_L = 3q$ and $n_R - m_R \neq 3q$   (or $n_L - m_L \neq 3q$ and $n_R - m_R = 3q'$; $q,q' \in \mathbb{Z}$), there is significant misalignment of the allowed momentum-energy manifolds corresponding to the two crystalline domains of graphene. Such periodic grain boundary structures (class II) are characterized by a transport gap

$$E_g = \hbar v_F \frac{2\pi}{3d} \approx \frac{1.38}{d[\text{nm}]} [\text{eV}],$$ (1)

which depends exclusively on the periodicity $d$. That is, class II grain boundaries perfectly reflect low-energy carriers in a remarkably large energy range $0.3-1.4$ eV for typical values of $d$ observed in experiments.

It is worth stressing that the momentum conservation rule does not involve any information about the details of the atomic structure in the interface region. Only easily accessible information about the grain boundary periodicity and its orientation with respect to the crystalline lattices of the two domains of graphene is sufficient to draw a conclusion about the possible presence of a transport gap and its magnitude. It also follows that all symmetric grain boundaries ($\theta_L = \theta_R$ and, hence, $n_L = n_R$ and $m_L = m_R$) belong to either class Ia or class Ib. More generally, all structures characterized by $d_{(n_L, m_L)} = d_{(n_R, m_R)}$, the so-called coincidence lattice sites grain boundaries, belong to one of these classes. Therefore, perfectly reflecting class II grain boundary defects are associated with lattice mismatch at the boundary line. It can be argued that non-decaying in-plane elastic strain would make such grain boundary structures less abundant or difficult to produce. However, one has to keep in mind that in 2D materials like graphene buckling in the third dimension provides an efficient mechanism for relieving in-plane compressive strain (see Ref. 19 and Supplementary Information for discussion).

In order to verify the results described above and gain further understanding of electronic transport across grain boundaries in graphene, we performed first-principles quantum transport calculations based on density functional theory and the non-equilibrium Green's function formalism (see Methods). Realistic atomic structures of two representative periodic grain boundaries constructed according to the dislocation model[16] are shown in Figure 3a,d.

The symmetric grain boundary shown in Figure 3a is defined by a pair of (2,1) matching vectors and, thus, it belongs to class Ib according to the above proposed scheme. This grain boundary structure is characterized by a very low formation energy of 3.4 eV/nm and by the smallest possible periodicity (0.65 nm) of all stress-free structures.[16] Such a structural motif has been proposed in the context of scanning tunnelling microscopy studies of grain boundaries in graphite.[6,20] The calculated electronic transmission as a function of transverse momentum and energy (Fig. 3b) agrees closely with the qualitative picture provided by the momentum conservation rule. Despite a high concentration of dislocations, such a structure is highly transparent with respect to charge carriers in a broad range of energies. The predicted total transmission probability through the grain boundary is ~0.8 for low-energy carriers, with values being slightly larger for the holes (Fig. 3c).

An example of a class II asymmetric grain boundary, formed by the (5,0) and (3,3) matching vectors, is shown in Figure 3d. Similar grain boundary configuration has recently been proposed in Ref. 21. The structure of this line defect contains three dislocations per repeat length of $d \approx 1.25$ nm. Although this structure involves a lattice mismatch $\Delta d/d = 3.8\%$, its formation energy of 5.0 eV/nm is still considerable smaller than the typical energies of bare graphene edges of ~10 eV/nm or more.[22] The system reveals an extraordinary large transport gap of 1.04 eV (Fig. 3e,f), in good agreement with $E_g = 1.1$ eV obtained from Eq. (1). In sharp contrast to the class Ib grain boundary discussed above, the transmission probabilities outside the transport gap are considerably lower than unity for both electrons and holes. Notably, the transmission probabilities do not respect electron-hole symmetry since the bipartite symmetry of graphene lattice is broken due to the presence of odd-membered rings in the interface region.

The possibility of introducing class II grain boundaries characterized by large transport gaps into graphene may find important practical applications. Due to the absence of band gap and the so-called Klein paradox, a counter-intuitive property originating from the charge conjugation symmetry of Dirac fermions, electrostatic barriers in graphene are inefficient.[23–25] While traditional electronics relies on tailoring potential energy profiles, the proposed approach exploits momentum mismatch as an instrument for controlling transport in graphene. However, the structural quality of grain boundaries is expected to be of crucial importance. We stress that the presence of periodicity-breaking disorder would inevitably introduce some leakage current in the predicted transport gap. However, we find that moderate amount randomly distributed short-range charge defects leads to only low conductance in the transport gap. In particular, our simulations of a model field effect transistor (FET) based on a reflecting grain boundary discussed above show that on/off current ratios above 1000 are achieved already at a concentration of short-range charge defects of 0.1 nm$^{-1}$ (see Supplementary Information for details). Such an on/off current ratio is considerably larger than the ones typically measured for top-gated single-layer graphene FETs (~5) and comparable to the highest values reported for dual-gate bilayer graphene devices.[26] Tailoring electronic transport by means of grain boundary engineering may pave a new road towards practical digital electronic devices based on graphene at truly nanometer scale.

## Methods

First-principles calculations have been performed using the density functional theory (DFT) approach implemented in the SIESTA code.[27] The generalized gradient approximation (GGA) exchange-correlation density functional[28] was used together with a double-$\zeta$ plus polarization basis set, norm-conserving pseudopotentials[29] and a mesh cutoff of 200 Ry. The 2D periodic computational models included two parallel equally spaced grain boundaries in a rectangular simulation supercell as described in Ref. 16. Both atomic coordinates and supercell dimensions were optimized using the conjugate-gradient algorithm and a 0.04 eV/Å maximum force convergence criterion. The electronic transport properties of the grain boundaries in graphene were calculated using the nonequilibrium Green's function formalism implemented in the TRANSIESTA code.[30] The quantum transmission was calculated in the zero-bias regime by including self-energies for the coupling of a 2 nm wide scattering region to the semi-infinite graphene leads.

**Ackowledgements** We are grateful to J. J. Palacios, C.-H. Park and D. Strubbe for their comments. This work was supported by National Science Foundation Grant No. DMR07-05941 and by the Director, Office of Science, Office of Basic Energy Sciences, Division of Materials Sciences and Engineering Division, U.S. Department of Energy under Contract No. DE-AC02-05CH11231. The structural parameters were determined using theoretical techniques and computer codes supported by NSF and the electronic transport calculations were carried out under the auspices of BES support. O.V.Y. acknowledges financial support of the Swiss National Science Foundation (grant No. PBELP2-123086). Computational resources have been provided by NSF through TeraGrid resources at NICS (Kraken) and by DOE at Lawrence Berkeley National Laboratory's NERSC facility.

**Author contributions** O.V.Y. proposed the project, carried out derivation, computations and analyses, and wrote the manuscript. S.G.L. directed the research, proposed analyses, interpreted results, and edited the manuscript.

**Additional information** Supplementary information accompanies this paper on www.nature.com/naturematerials. Reprints and permissions information is available online at http://npg.nature.com/reprintsandpermissions. Correspondence and requests for materials should be addressed to O.V.Y. (yazyev@civet.berkeley.edu) or S.G.L (sglouie@berkeley.edu).

**Figure 1 | Structure of grain boundaries in graphene. a**, An example of tilt grain boundary in graphene separating two crystalline domains rotated by $\theta = \theta_L + \theta_R = 8.2° + 30.0° = 38.2°$ with respect to each other. We use the convention for misorientation angles introduced in Ref. 16. The repeat vector $\vec{d}$ of the grain boundary structure is defined by the matching vectors (5,3) and (7,0) in the left and right domains, respectively. A possible atomic structure of the interface region involves three elementary dislocation dipoles (pentagon-heptagon pairs) per repeat cell. **b**, Effective rotation of the hexagonal Brillouin zone of graphene experienced by charge carriers crossing the (5,3)|(7,0) grain boundary.

**Figure 2 | Grain boundaries in graphene − two distinct transport behaviours. a**, Scheme illustrating the mapping of the band structures of the two crystalline domains of graphene onto the 1D mini-BZ of the periodic grain boundary structure. The BZs of the two graphene domains (blue and red hexagons) rotated by angles $\theta_L$ and $\theta_R$ respectively, are folded along the dotted lines $k_\parallel = (2n+1)\pi/d$ $(n \in \mathbb{Z})$, and projected onto the 1D mini-BZ of the periodic grain boundary (thick line). The actual construction corresponds to the (5,3)|(7,0) grain boundary shown in Figure 1a. **b**, Position of the 2D BZ corners $K$ and $K'$ with respect to the solid lines $k_\parallel = 2n\pi/d$ for $n - m = 3q$ and $n - m \neq 3q$ $(q \in \mathbb{Z})$. **c**, Shaded areas correspond to momentum-energy pairs for which conductance channels exist in the two crystalline domains. In the case of class Ia or class Ib grain boundaries, two and one conductance channels, respectively, are available for allowed values of $k_\parallel$ at low energies. Denser shading in the case of class Ib grain boundaries at higher energies corresponds to two conductance channels. No selection by momentum (i.e., mismatch of $k_\parallel$ for states at a given energy) takes place for these two classes since identical areas of available conductance channels are superimposed. Transmission

through class II grain boundaries is possible only in regions where the two colours overlap (shown in magenta), opening a transport gap $E_g$.

**Figure 3 | Electronic transport through grain boundaries in graphene from first principles**. **a**, Atomic structure of the (2,1)|(2,1) ($\theta = 21.8°$) class Ib grain boundary. **b**, Transmission probability through the (2,1)|(2,1) grain boundary as a function of transverse momentum $k_\parallel$ and energy $E$. **c**, Zero-bias total transmission per unit length through the (2,1)|(2,1) grain boundary as a function of energy (solid line) compared to the one of ideal graphene (dashed line). **d**, Atomic structure of the (5,0)|(3,3) ($\theta = 30.0°$) class II grain boundary. **e,f**, Corresponding $k_\parallel$-resolved and total charge carrier transmissions through the (5,0)|(3,3) class II grain boundary.

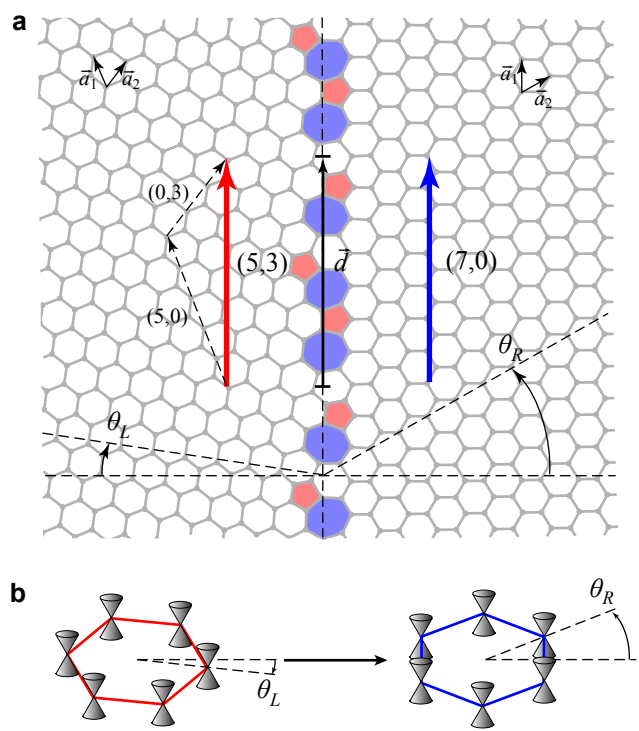

Figure 1
Yazyev & Louie (2010)

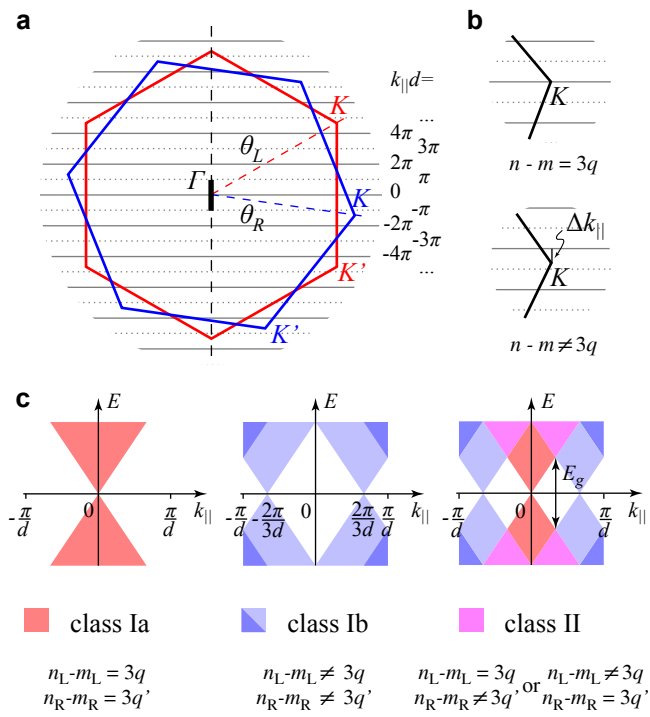

**a**

$k_\parallel d =$

$\cdots$
$4\pi$
$3\pi$
$2\pi$
$\pi$
$0$
$-\pi$
$-2\pi$
$-3\pi$
$-4\pi$
$\cdots$

$K$
$K$
$\Gamma$
$\theta_L$
$\tilde\theta_R$
$K'$
$K'$

**b**

$K$

$n - m = 3q$

$\tilde{\Delta k}_\parallel$

$K$

$n - m \neq 3q$

**c**

$E$

$-\frac{\pi}{d}$   $0$   $\frac{\pi}{d}$   $k_\parallel$

$E$

$-\frac{\pi}{d}$  $-\frac{2\pi}{3d}$   $0$   $\frac{2\pi}{3d}$  $\frac{\pi}{d}$   $k_\parallel$

$E$

$E_g$

$-\frac{\pi}{d}$   $0$   $\frac{\pi}{d}$   $k_\parallel$

■ class Ia          ■ class Ib          ■ class II

$n_L\text{-}m_L = 3q$          $n_L\text{-}m_L \neq 3q$          $n_L\text{-}m_L = 3q,$  $n_L\text{-}m_L \neq 3q$
$n_R\text{-}m_R = 3q'$          $n_R\text{-}m_R \neq 3q'$          $n_R\text{-}m_R \neq 3q',$ or $n_R\text{-}m_R = 3q'$

Figure 2
Yazyev & Louie (2010)

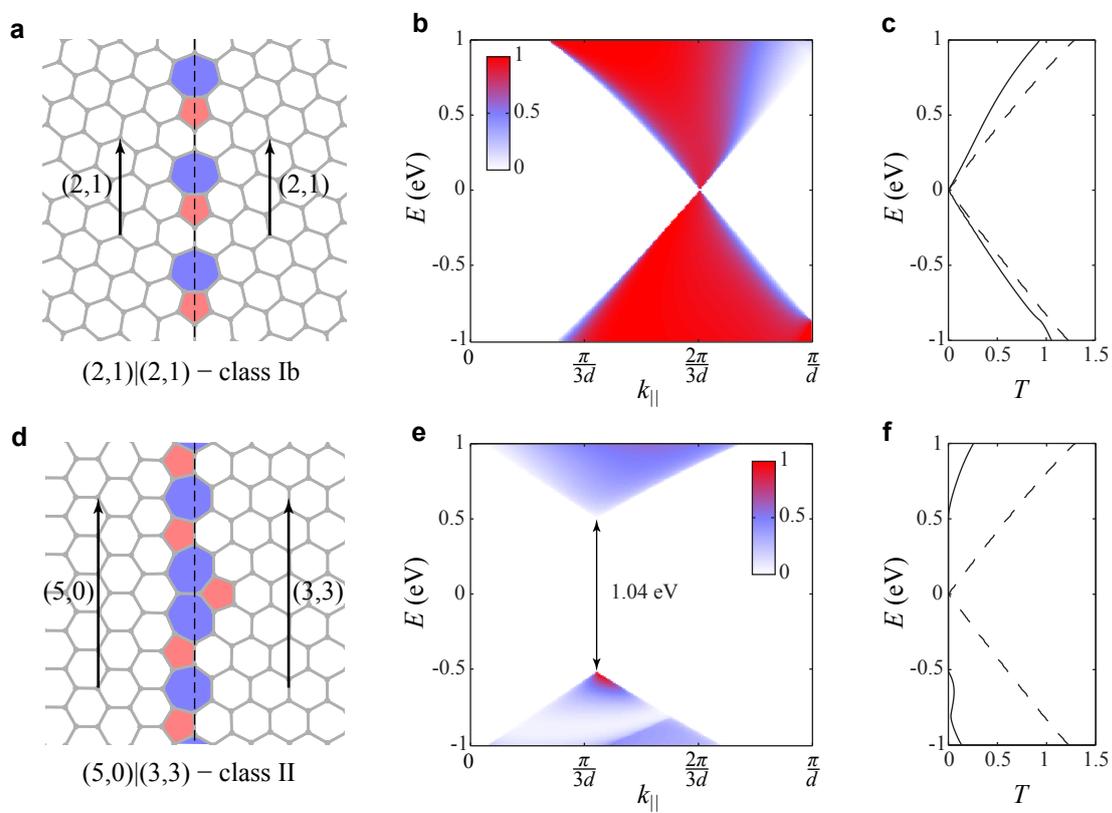

**a**

(2,1)    (2,1)

(2,1)|(2,1) − class Ib

**b**

$E$ (eV)

$\frac{\pi}{3d}$    $\frac{2\pi}{3d}$    $\frac{\pi}{d}$

$k_{||}$

**c**

$E$ (eV)

$T$

**d**

(5,0)    (3,3)

(5,0)|(3,3) − class II

**e**

$E$ (eV)

1.04 eV

$\frac{\pi}{3d}$    $\frac{2\pi}{3d}$    $\frac{\pi}{d}$

$k_{||}$

**f**

$E$ (eV)

$T$

Figure 3
Yazyev & Louie (2010)

# Supplementary Information for
# "Electronic transport in polycrystalline graphene"

Oleg V. Yazyev and Steven G. Louie

*Department of Physics, University of California,*

*Berkeley, California 94720, USA and*

*Materials Sciences Division, Lawrence Berkeley*

*National Laboratory, Berkeley, California 94720, USA*

**CONTENTS:**

# 1. Bicrystallographic properties of grain boundaries in graphene and lattice-mismatch strain relief.

The length of translational vector $(n, m) = n\vec{a}_1 + m\vec{a}_2$ on the hexagonal lattice of graphene

$$d_{(n,m)} = a_0\sqrt{n^2 + nm + m^2}, \tag{1}$$

where $a_0 = 0.246$ nm is the length of unit vectors $\vec{a}_{1,2}$. Since $(d_{(n,m)}/a_0)^2 = n^2 + nm + m^2 = (n - m)^2 + 3nm$, the following relation holds:

$$\left(\frac{d_{(n,m)}}{a_0}\right)^2 = 3p \quad \text{iff} \quad n - m = 3q \quad (p \in \mathbb{N}, q \in \mathbb{Z}). \tag{2}$$

That is, for class II grain boundaries characterized by $n_L - m_L = 3q$ and $n_R - m_R \neq 3q$ (or $n_L - m_L \neq 3q$ and $n_R - m_R = 3q$) the exact commensurability condition $d_{(n_L,m_L)} = d_{(n_R,m_R)}$ ($d_L = d_R$ below) is never satisfied. In other words, a finite lattice mismatch $\Delta d/d = (d_R - d_L)/d_L$ ($d_R > d_L$) is always present in the periodic class II grain boundary structures.

Similarly to bulk 3D materials, in the case of 2D membranes constrained to plane such lattice mismatch leads to residual tensile and compressive strain fields as one moves away from the interface in the right and left domains, respectively. The residual elastic energy per unit area of membrane converges to

$$E_S = \frac{1}{2}Y\epsilon_x^2, \tag{3}$$

where $Y \sim 1$ TPa [S1] is the Young's modulus and $\epsilon_x = \Delta d/(2d)$ ($\epsilon_x = -\Delta d/(2d)$) is the in-plane strain in the left (right) domain. Such behavior would 'penalize' lattice mismatched grain boundaries making them less abundant in polycrystalline samples or difficult to produce on purpose.

However, free-standing or weakly bound 2D membranes behave in a remarkably different way [S2-S5]. In this case, the in-plane compressive strain is efficiently relieved by the out-of-plane deformation as schematically illustrated in Figure S1a. If one assumes that in-plane strain in the right domain is fully relieved by developing periodic ripples which are described by the out-of-plane displacement

$$\xi(x) = \frac{A}{2}\sin\frac{2\pi x}{\lambda}, \tag{4}$$

characterized by wavelength $\lambda$ and amplitude

$$A = \frac{2\lambda}{\pi}\sqrt{\frac{\Delta d}{d}}, \tag{5}$$

then, the average bending energy per unit area of such rippled domain is

$$\langle E_b \rangle = \frac{1}{2}\kappa \langle (\xi''_{xx})^2 \rangle = 4\kappa \frac{\pi^2}{\lambda^2}\frac{\Delta d}{d}, \tag{6}$$

where $\kappa \sim 1$ eV [S6,S7] is the bending rigidity of graphene. Average bending energy $\langle E_b \rangle$ can assume arbitrarily small values in the long-wavelength regime. An intuitive physical picture of this phenomenon is the screening of in-plane elastic fields by the out-of-plane deformation. Thus, the formation energies of the grain boundaries in graphene characterized by reasonably small lattice mismatches are expected to be comparable to the lattice-matched ones.

Below, we further discuss bicrystallographic properties of grain boundaries in graphene. The 'treasure map' diagram shown in Figure S1b locates all possible lattice-matched class Ia and Ib grain boundaries according to their misorientation angle $\theta$ and periodicity $d < 5$ nm. Most of these grain boundary configurations are symmetric. In particular, the configu- rations investigated in Ref. S8 correspond to the lowest branch labeled on the diagram. The $(2,1)|(2,1)$ configuration used for illustrating the transport behavior of class Ib grain boundaries in the present article corresponds to the LAGB I configuration of Ref. S8. An- other remarkable configuration characterized by the $(3,2)$ matching vectors corresponds to the LAGB II structure which has been determined as the lowest-energy large-angle grain boundary in graphene [S8]. Asymmetric lattice-matched grain boundaries are also present on the diagram (filled circles in Fig. S1b). The smallest possible periodicity $d = 1.72$ nm is realized in the $(5,3)|(7,0)$ configuration which is used as a generic example in the main discussion of the present work. Interestingly, asymmetric lattice-matched grain boundaries with periodicities $d < 5$ nm are realized only in the large-angle regime for some specific 'magic' values $\theta = 21.8°$, $27.8°$, $32.2°$, $38.2°$ and $46.8°$.

Figure S1c shows all possible class II grain boundaries characterized by lattice mismatches $0.01 \leq \Delta d/d < 0.04$ (open circles) and $\Delta d/d < 0.01$ (filled circles). Mismatches $\Delta d/d < 0.04$ are realized for periodicities $d \gtrsim 1$ nm while $\Delta d/d < 0.01$ requires $d \gtrsim 2$ nm. More generally, we find that that the minimum $d$ possible for a give value of $\Delta d/d$ is described by $d = \omega(\Delta d/d)^{-1/2}$ ($\omega \approx 0.174$ nm) as shown in Figure S1d. The $(5,0)|(3,3)$ configuration used for illustrating the transport behavior of class II grain boundaries is labeled in Figures S1c and S1d.

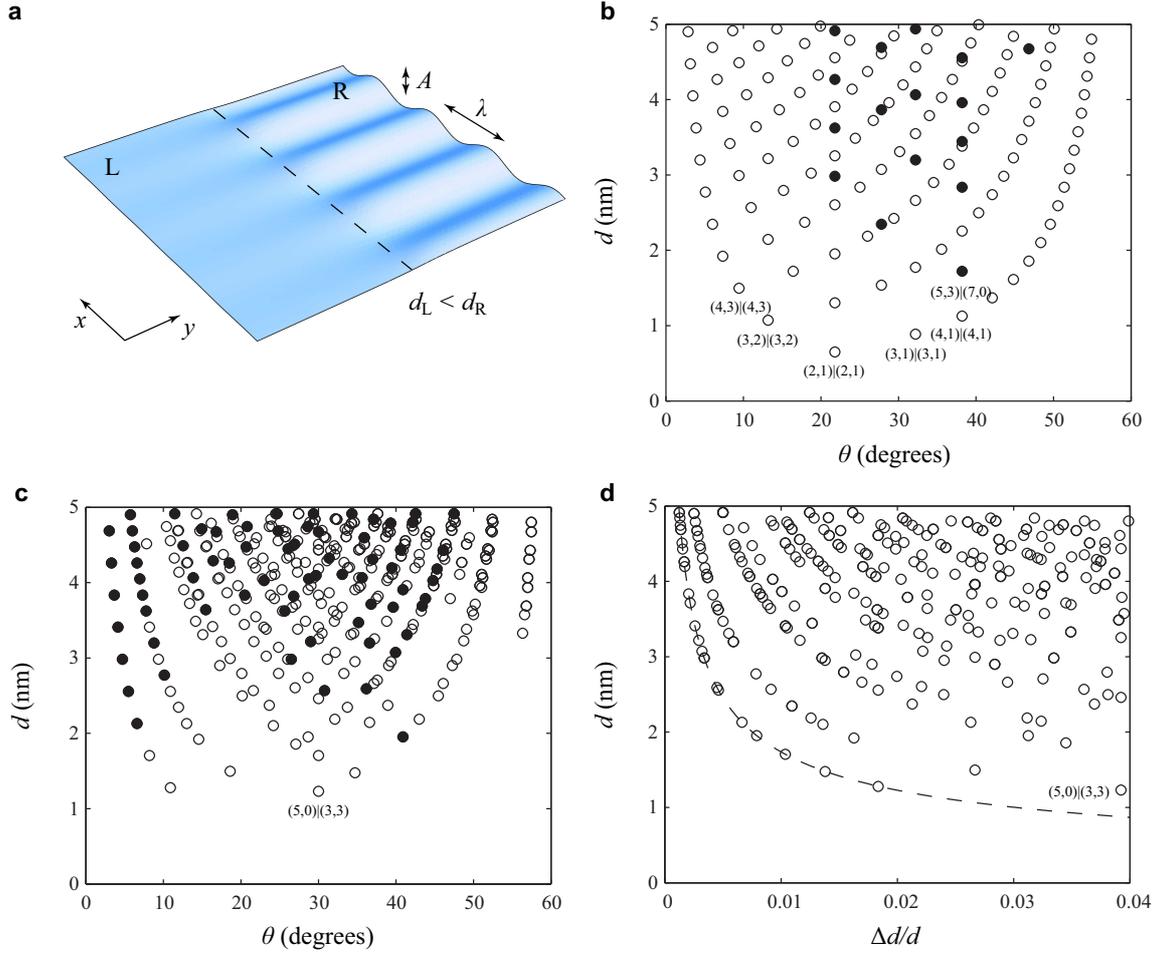

FIG. S1: (a) Schematic illustration of the lattice-mismatch strain relief at grain boundaries in 2D membranes. Right domain is characterized by the longer matching vector, $d_\mathrm{R} > d_\mathrm{L}$. Dashed line shows the boundary location. (b) Symmetric (open circles) and asymmetric (filled circles) class Ia and class Ib grain boundaries shown on a 'treasure map' diagram according to their misorientation angle $\theta$ and periodicity $d$. (c) Class II grain boundaries characterized by lattice mismatches $0.01 \leq \Delta d/d < 0.04$ (open circles) and $\Delta d/d < 0.01$ (filled circles). (d) Class II grain boundaries on a diagram showing periodicity $d$ vs. lattice mismatch $\Delta d/d$. The dashed line $d = \omega(\Delta d/d)^{-1/2}$ ($\omega \approx 0.174$ nm) shows the lower limit of $d$ at given $\Delta d/d$. Selected grain boundary configurations are labeled in panels (b)-(d).

## 2. Effect of disorder on charge transport across class II grain boundaries

In order to study the effect of disorder onto the conductance of class II grain boundaries we carry out quantum transport calculations with impurities randomly distributed in a model with supercell dimension many times larger than the grain boundary periodicity. Systematic first-principles transport calculations on such extended models cannot be performed due to their computational cost. To overcome this limitation we investigate transport properties of large models within an approach based on the tight-binding model Hamiltonian

$$\mathcal{H} = -t \sum_{\langle i,j \rangle} [c_i^\dagger c_j + \text{h.c.}], \tag{7}$$

where $c_i$ ($c_i^\dagger$) annihilates (creates) an electron at site $i$ and $\langle i,j \rangle$ stands for pairs of nearest-neighbor atoms. The hopping integral $t = 2.66$ eV is assumed to be constant. Transmission as a function of momentum $k_{||}$ and energy $E$

$$T(k_{||}, E) = \text{Tr} \left[ \Gamma_\text{L}(k_{||}, E) G_\text{S}^\dagger(k_{||}, E) \Gamma_\text{R}(k_{||}, E) G_\text{S}(k_{||}, E) \right] \tag{8}$$

is evaluated from the Green's function of the grain boundary scattering region

$$G_\text{S}(k_{||}, E) = \left[ E^+ I - \mathcal{H}_\text{S} - \Sigma_\text{L}(k_{||}, E) - \Sigma_\text{R}(k_{||}, E) \right]^{-1} \tag{9}$$

and the coupling matrices

$$\Gamma_\text{L(R)}(k_{||}, E) = i \left( \Sigma_\text{L(R)}(k_{||}, E) - \Sigma_\text{L(R)}^\dagger(k_{||}, E) \right) \tag{10}$$

for the left lead (the right) lead. Here, $E^+ = E + i\eta$ ($\eta \to 0^+$), $I$ is a unit matrix, $\mathcal{H}_\text{S}$ is the Hamiltonian of the scattering region and $\Sigma_\text{L(R)}(k_{||}, E)$ are the self-energies coupling to the ideal graphene leads [S9].

The applicability of tight-binding Green's function approach is verified by comparing the $T(k_{||}, E)$ maps with the corresponding first-principles results for the models of class Ib and class II grain boundaries discussed in the main text (Fig. S2). Clearly, the results for the class Ib grain boundary show very good quantitative agreement. The transport gap and all salient features in the conductance map of the class II grain boundary are also reproduced although the region of high transmission for the holes is somewhat exaggerated in the tight-binding calculations. Good agreement between the model Hamiltonian and first-principles approaches allows us to conclude that transmission probabilities through the grain boundaries in graphene are governed primarily by the covalent bond connectivity at the interface.

It is important to comment on the electron-hole symmetry of the transmission maps calculated using the tight-binding model Hamiltonian. The limits of the regions of finite transmission probabilities are symmetric with respect to the inversion of the energy scale since the momentum conservation law depends exclusively on the mutual orientation of ideal graphene domains. Within the nearest-neighbor tight-binding Hamiltonian (7) the band structure of the bipartite lattice of ideal graphene exhibits the property of electron-hole symmetry. However, the bipartite symmetry of the graphene lattice is locally broken at the grain boundary interface. As a result, the transmission probabilities through the grain boundary need not respect electron-hole symmetry.

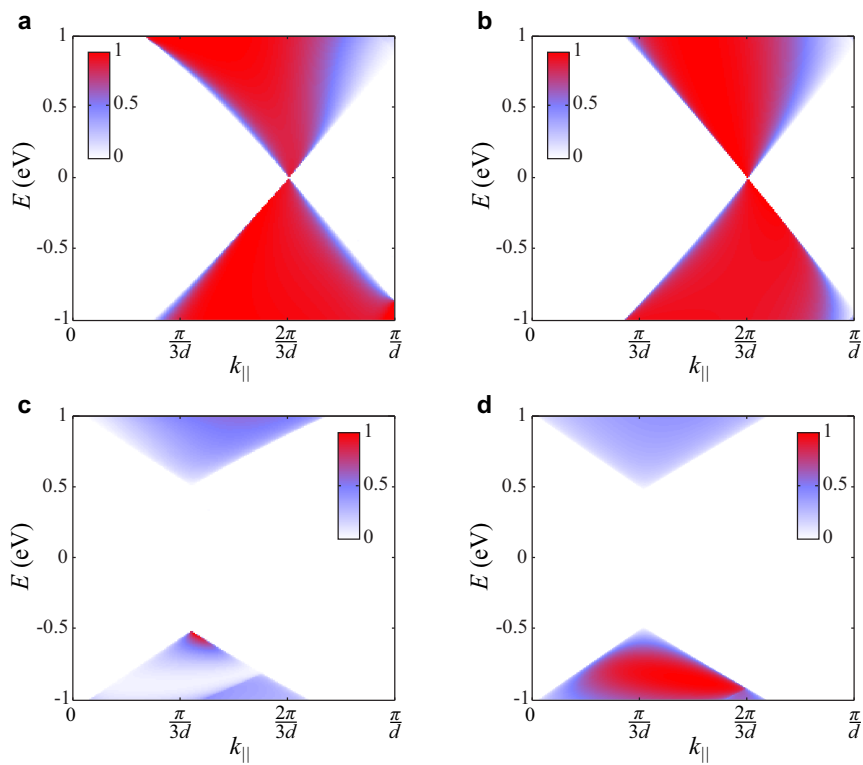

FIG. S2: Transmission probability through the (2,1)|(2,1) ($\theta = 21.8°$) class Ib grain boundary as a function of transverse momentum $k_{||}$ and energy $E$ calculated using (a) first principles and (b) tight-binding Green's function approaches. Transmission probability through the (5,0)|(3,3) ($\theta = 30.0°$) class II grain boundary as a function of transverse momentum $k_{||}$ and energy $E$ calculated using (c) first-principles and (d) tight-binding approaches. The value of hopping parameter $t = 2.66$ eV is assumed in the tight-binding calculations.

Transport properties of the disordered class II grain boundary are investigated using a supercell model involving 9 repeat units of the (5,0)|(3,3) interface. The resulting supercell periodicity is ∼11 nm. Different disorder realizations are modeled by random placing of 1–10 short-ranged impurities at the interface. This corresponds to the range of impurity concentrations $c = 0.1 − 1$ nm$^{-1}$. The impurities are introduced into the scattering region Hamiltonian through

$$\mathcal{H}'_{\text{S}} = \sum_i \epsilon_i c_i^\dagger c_i, \tag{11}$$

where an on-site impurity potential $\epsilon_i = -0.1t$ is assumed in the present calculations. For each impurity concentration the results are averaged over ensembles of 200 different realizations of disorder. An example of zero-bias transmission as a function of energy $E$ for 10 different configurations at impurity concentration $c = 0.9$ nm$^{-1}$ is shown in Figure S3a (solid lines). The presence of disorder induces weak conductance in the transport gap around $E = 0.1t$. The effect of disorder on conductance is very small outside the transport gap.

We further investigate the effect of disorder onto the on/off current ratios in a model field-effect transistor (FET) based on the considered class II grain boundary. The on/off current ratios are calculated using the following expression for the current at gate voltage $V_g$ and bias voltage $V_b$

$$I(V_g, V_b) = \frac{2e^2}{h} \int_{eV_g - eV_b/2}^{eV_g + eV_b/2} \int_{k_{||}} T(k_{||}, E) dk_{||} dE. \tag{12}$$

Due to the large transverse dimension of the supercell momentum is sampled only at $k_{||} = 0$. For the OFF state of the model transistor we choose the gate voltage $V_g = 0.1t/e$ which places the chemical potential in the region of the disorder-induced conductance. The value of gate voltage $V_g = 0.4t/e$ for the ON state and a bias voltage $V_b = 0.2t/e$ is assumed for both the ON and OFF state. For the present case of weak short-range disorder, the calculated on/off current ratios (Fig. S3b) exhibit the $1/c$-dependence on the impurity concentration $c$ showing that $I_{\text{on}}/I_{\text{off}}$ values over 1000 are achieved for impurity concentrations $c < 0.1$ nm$^{-1}$.

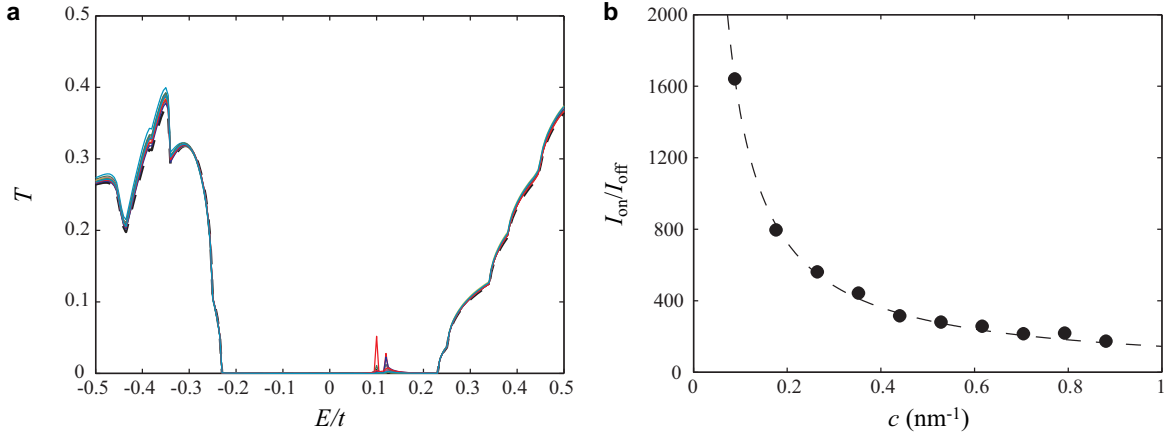

FIG. S3: (a) Electronic transmission as a function of energy at zero bias across the disordered (5,0)|(3,3) grain boundary for 10 different random realizations of disorder at a concentration of impurities $c = 0.9$ nm$^{-1}$ (solid lines). Transmission in the absence of disorder is shown for reference (dashed line). (b) Calculated on/off current ratio $I_{on}/I_{off}$ as a function of impurity concentration $c$ ($V_g(\text{on}) = 0.4t/e$, $V_g(\text{off}) = 0.1t/e$, $V_b = 0.2t/e$). Dashed line shows a $vc^{-1}$ fit ($v = 125$ nm$^{-1}$).